\documentclass[sigconf]{acmart}
\usepackage{multirow}
\usepackage{balance}
\AtBeginDocument{%
  }

\copyrightyear{2025}
\acmYear{2025}
\setcopyright{cc}
\setcctype{by}
\acmConference[Preprint]{}
\acmISBN{}

\begin{document}

\title{Towards Efficient Hypergraph and Multi-LLM Agent Recommender Systems}

\author{Tendai Mukande}
\affiliation{%
\institution{Research Ireland ML-LABS}
  \institution{Dublin City University}
  \city{Dublin}
  \country{Ireland}}
\email{tendai.mukande2@mail.dcu.ie}
\author{Esraa Ali}
\affiliation{%
  \institution{ADAPT Centre} 
 \institution{Dublin City University}
  \city{Dublin}
  \country{Ireland}}
\email{abdelmoe@tcd.ie}

\author{Annalina Caputo}
\affiliation{%
 \institution{School of Computing}
  \institution{Dublin City University}
  \city{Dublin}
  \country{Ireland}}
\email{annalina.caputo@dcu.ie}

\author{Ruihai Dong}
\affiliation{%
  \institution{Insight Research Ireland Centre for Data Analytics}
  \institution{University College Dublin}
  \city{Dublin}
  \country{Ireland}}
\email{ruihai.dong@ucd.ie}

\author{Noel O'Connor}
\affiliation{%
\institution{Insight Research Ireland Centre for Data Analytics}
  \institution{Dublin City University}
  \city{Dublin}
  \country{Ireland}}
\email{Noel.OConnor@dcu.ie}


\begin{abstract}
Recommender Systems (RSs) have become the cornerstone of various applications such as e-commerce and social media platforms. The evolution of RSs is paramount in the digital era, in which personalised user experience is tailored to the user's preferences. Large Language Models (LLMs) have sparked a new paradigm - generative retrieval and recommendation. Despite their potential, generative RS methods face issues such as \textbf{hallucination}, which degrades the recommendation performance, and \textbf{high computational cost} in practical scenarios. To address these issues, we introduce \textbf{HGLMRec}, a novel multi-LLM agent-based RS model that incorporates a hypergraph encoder designed to capture
complex relationships between users and items. The HGLMRec model retrieves only the relevant tokens during inference, reducing computational overhead while enriching the retrieval context. Experimental results show performance improvement by HGLMRec against state-of-the-art baselines at lower computational cost.
\end{abstract}




\maketitle

\section{Introduction}
In real-world recommendation scenarios, user preferences evolve continuously over time, leading to complex and dynamic interaction patterns \cite{liu2024multi}. Static representation learning models \cite{fang2021session,kang2018self, sun2019bert4rec} often struggle to capture these dynamics as they assume a static structure and ignore the time-dependent nature of user-item interactions \cite{zhao2024dynllm}. With the recent surge in LLMs, a new paradigm in RSs has emerged that combines information retrieval with LLMs to produce contextually relevant recommendations \cite{chen2024large, li2024large, mysore2023large,sileo2022zero}. Generative recommendation models have shown benefits, such as semantic understanding and interactive reasoning, which can improve the relevance and quality of recommendations by generating output that aligns with user preferences \cite{li2024matching, wang2023generative}. 

Despite these advances, most of the existing LLM-based RS approaches face two major limitations. Firstly,\textbf{ hallucination}, where the model generates inaccurate or misleading recommendations, can compromise the reliability of the system \cite{he2023large, huang2023recommender,ji2024genrec}. Secondly, the \textbf{high computational cost}, resulting from the need to search through large vocabularies or fine-tune LLMs on domain-specific data, makes these methods impractical for real-time or large-scale deployment \cite{du2024large,wei2022chain,yao2024tree}. Although pretraining or fine-tuning LLMs in recommendation-specific datasets can improve performance, these strategies require substantial computational resources \cite{li2024survey, zhang2024llm, fatemi2024test}, domain expertise, and large volumes of high-quality data, further complicating real-world implementation \cite{zhou2024survey}. Consequently, efficient recommendation models are needed that can adapt to evolving user preferences and dynamic interaction patterns while maintaining high accuracy \cite{harte2023leveraging, xi2023towards}. 

\begin{figure}[!ht]
   \centering
\includegraphics[width = \columnwidth]{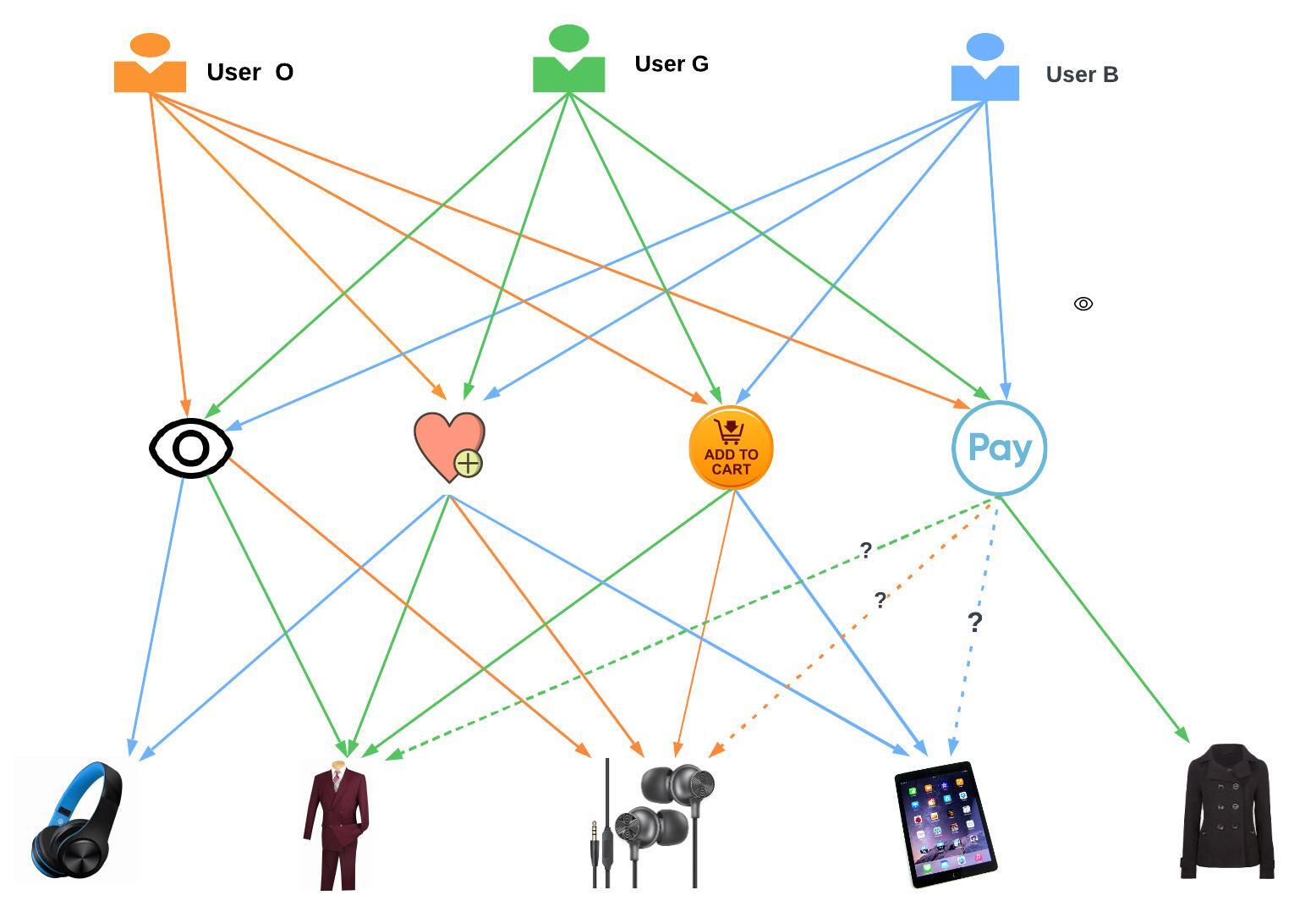}
    \Description{A real-world e-commerce recommendation scenario involving multiple user behaviour types.}
    \caption{An example of a real-world e-commerce recommendation scenario involving multiple user behaviour types. This scenario is best represented by a hypergraph, allowing modelling of higher-order, multi-way relationships. \textbf{Nodes:} users/items. \textbf{Hyperedges:} view, add-to-cart, purchase, etc}
    \label{c1_fig:MBR}
    
\end{figure}

Motivated by these challenges, we explore whether hypergraph representation learning can be harnessed to improve recommendation performance in dynamic, multi-behaviour scenarios. As illustrated in Figure \ref{c1_fig:MBR}, hypergraphs allow modelling of higher-order user-item interactions, unlike bipartite graphs that are limited to pairwise interactions \cite{li2023hypergraph}. We propose \textbf{HGLMRec}, a novel framework that integrates an HGNN encoder with an MoA architecture. The central idea of HGLMRec is to enhance modelling of user-item interactions, allowing the model to capture higher-order dependencies across multiple behaviours. Hyperedges in this representation connect a user with multiple items and behavioural types, generating dense token embeddings that encode local and global preference patterns \cite{bai2021hypergraph}. These embeddings are then processed by the MoA framework, which employs multiple specialised agents to refine recommendations. 

Our \textbf{contributions} are as follows. We apply an HGNN encoder to capture complex, higher-order interactions that static pairwise GNNs often miss \cite{kim2020hypergraph}, and integrate embeddings across multiple behaviours to preserve the richness of user-item interactions. 
Inspired by the work in \cite{wang2024mixture}, we pass the HGNN and user prompt tokens through an MoA framework that learns dynamic patterns in user behaviour. We hypothesise that this combination allows HGLMRec to achieve expressive representation learning while remaining computationally efficient, making it suitable for large-scale deployment. Experiments on three real-world datasets demonstrate that our proposed model, HGLMRec, outperforms state-of-the-art baselines, including several models based on LLMs, at lower computational cost.

\begin{figure*}
    \centering\includegraphics[width=\textwidth]{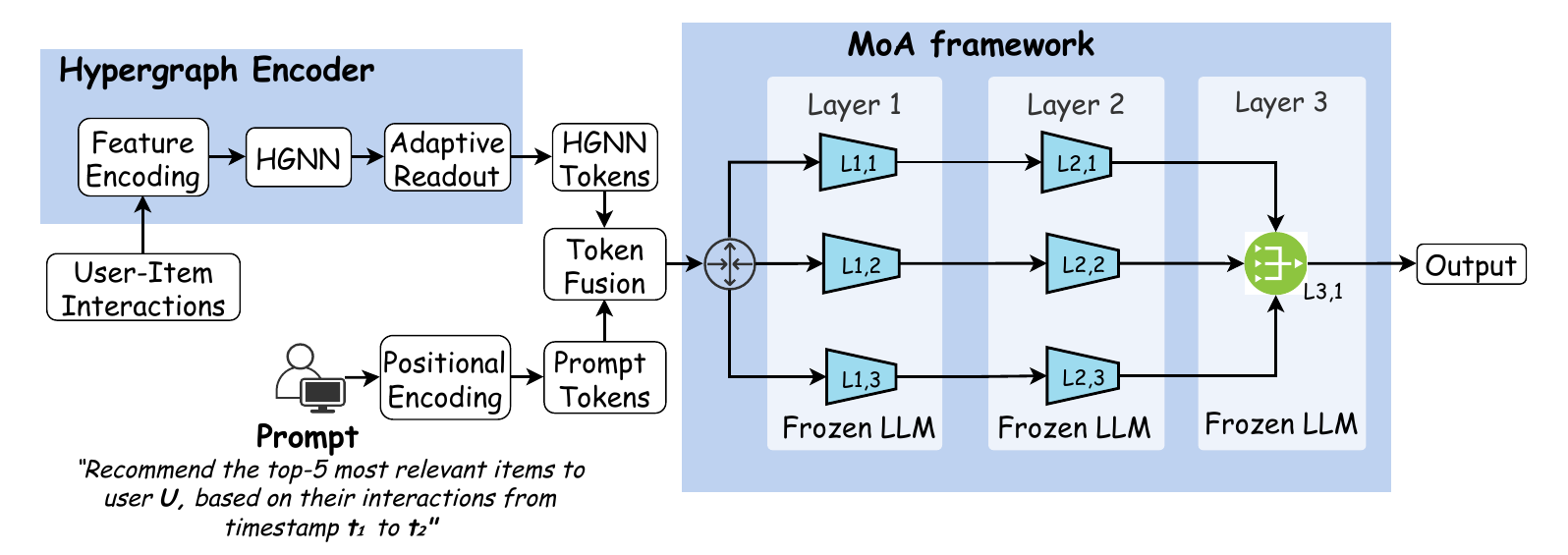}
    \Description{Illustration of the HGLMRec model architecture.}
    \caption{End-to-end architecture of HGLMRec. The model processes user-item interactions through an encoder, fuses graph tokens with task prompts, and refines recommendations via a 3-layer MoA framework. These agents leverage interactions captured by the encoder . The final MoA layer (L3,1) aggregates information from the intermediate LLM agents.}
    \label{c7_fig:HGLM}
\end{figure*} 

\section{Related Work}

\paragraph{Graph and LLM Integration.}
The integration of graph structures with LLMs has gained attention due to their complementary strengths \cite{zhang2023graph, li2024graph}. Graphs explicitly encode relationships between entities, offering a structured representation that enhances reasoning and reduces the risk of hallucination in LLMs \cite{fatemi2023talk}. LLMs, on the other hand, excel in understanding and generating unstructured text, but lack mechanisms for directly incorporating graph-based knowledge \cite{xi2023towards}. Previous studies have explored the use of knowledge graphs with LLM for tasks such as node classification, link prediction and question answering \cite{mai2023hypergraph}. These approaches often rely on fine-tuned LLM instruction to align graph-based reasoning with task objectives \cite{zhao2024dynllm}. However, they are limited by their dependence on static graphs and their inability to take advantage of the rich relational information captured by hypergraphs \cite{luo2020dynamic}. In addressing this issue, Fatemi et al. \cite{fatemi2023talk} propose  GraphToken, a parameter-efficient technique that augments textual prompts with learned tokens that represent graph structures, to encode structured data for LLM. Instead of fine-tuning the entire LLM, GraphToken trains a small set of additional parameters that encode the structure of the input data, allowing the LLM to better understand the data without extensive fine-tuning  \cite{perozzi2024let}.  In this line of research, Perozzi et al. \cite{perozzi2024let} introduce the
GraphToken model, which encodes structured data for LLMs, generating continuous token representations that encode the structure and attributes of graph data, bypassing the limitations of traditional lexical tokenisation \cite{fatemi2023talk}. Inspired by their work, our HGLMRec extends this concept to RS scenarios.

\section{Methodology}
 The architecture of HGLMRec, illustrated in Figure \ref{c7_fig:HGLM}, consists of three interconnected stages. The first stage, hypergraph encoding, constructs hypergraph representations of multi-behaviour interactions and generates initial token embeddings. These embeddings are then passed through the token fusion stage, which integrates information across behaviours to produce unified representations that encode both fine-grained and global patterns of user-item interactions. Finally, hierarchical MoA processing refines these fused embeddings through multiple specialised agents, with dynamic weighting to produce the final recommendation scores.

HGLMRec is designed to overcome the limitations of static HGNNs and effectively models complex multi-behaviour dependencies \cite{luo2020dynamic, zhao2024dynllm}. The hierarchical MoA framework refines embeddings through specialised agents, reducing hallucinations commonly observed in generative LLM-based approaches and improving reliability of the recommendations. Moreover, the lightweight design of both the hypergraph encoder and the MoA components allows efficient training and inference, offering a computationally feasible alternative to large-scale LLM-based recommendation models. Figure \ref{c7_fig:HGLM} illustrates the architecture, which consists of three key stages: (1) Hypergraph Encoding, (2) Token Fusion, and (3) Hierarchical MoA Processing.

\subsection{Hypergraph Encoder}
\label{subsec:hgnn}

Given a hypergraph \(\mathcal{G} = (\mathcal{V}, \mathcal{E})\), where the node set \(\mathcal{V} = U \cup I\) consists of users \(U\) and items \(I\), and the hyperedges \(\mathcal{E}\) capture multi-behaviour interactions such as views, cart additions and purchases, the hypergraph encoder produces compact user-item representations through the following steps:

\paragraph{Feature Initialisation.}  
Each node \(v \in \mathcal{V}\) is initialised with a learnable embedding vector:
\begin{equation}
\mathbf{h}_v^{(0)} = \begin{cases}
    \mathbf{E}_u[v] \in \mathbb{R}^d & \text{if } v \in U, \\
    \mathbf{E}_i[v] \in \mathbb{R}^d & \text{if } v \in I,
\end{cases}
\label{eq:feature_init}
\end{equation}
where \(\mathbf{E}_u\) and \(\mathbf{E}_i\) are embedding lookup tables for users and items, respectively, and \(d\) is the embedding dimension.

\paragraph{Hypergraph Convolution.}  
To capture higher-order user-item interactions, we apply two layers of hypergraph convolution. At each layer \(l \in \{0,1\}\), the node features are updated by aggregating the normalised messages from all hyperedges:
\begin{equation}
\mathbf{h}_v^{(l+1)} = \text{LayerNorm} \left( \sigma \left( \sum_{e \in \mathcal{E}(v)} \frac{1}{|e|} \sum_{u \in e} \mathbf{h}_u^{(l)} \mathbf{W}^{(l)} \right) \right),
\label{eq:hgnn_layer}
\end{equation}
where \(\mathcal{E}(v)\) is the set of hyperedges containing node \(v\), \(|e|\) is the size of hyperedge \(e\), \(\mathbf{W}^{(l)} \in \mathbb{R}^{d \times d}\) is a learnable weight matrix, \(\sigma\) is the ReLU activation function, and \textit{LayerNorm} stabilises  training.

\paragraph{Adaptive Readout.} To improve the representation learning of user-item interactions from the HGNN module, we apply adaptive readout \cite{buterez2022graph}, a  function that aggregates the HGNN embeddings \cite{xu2018powerful}. In HGLMRec, readout is applied using an attention-weighted grouping (Equation \ref{eq_adaptive}) that retains expressive hypergraph summaries tailored to the particular interactions for the particular iteraction.

\paragraph{Token Generation.}  
After two convolutional layers, node embeddings \(\{\mathbf{h}_v^{(l)}\}_{v\in\mathcal{V}}\) are pooled by leveraging  \emph{adaptive readout} to form compact graph tokens:
\begin{equation}
    \alpha_v = \frac{\exp\left(\mathbf{a}^\top \tanh\left(\mathbf{W}_a\,\mathbf{h}_v^{(l)}\right)\right)}{\sum_{u\in\mathcal{V}} \exp\left(\mathbf{a}^\top \tanh\left(\mathbf{W}_a\,\mathbf{h}_u^{(l)}\right)\right)}, \quad
    \mathbf{G} = \mathrm{MLP}\Bigl(\sum_{v\in\mathcal{V}} \alpha_v\,\mathbf{h}_v^{(l)}\Bigr)
\label{eq_adaptive}
\end{equation}
where \(\alpha_v\) are attention weights learned via \(\mathbf{W}_a\) and \(\mathbf{a}\), and the MLP ensures flexible mapping of aggregated features.

\subsection{Token Fusion}
\label{subsec:fusion}

To align graph-based and prompt signals, HGLMRec fuses \(\mathbf{G}\) with a tokenised task prompt. We apply token fusion by concatenation \cite{li2024graph} to the model that combines structured graph tokens with the recommendation task prompt. Specifically, a prompt such as ``\textit{Recommend the top-5 most relevant items for user U, based on their interactions from timestamp t2 to t1}'', is tokenised, which converts the text into a sequence of token embeddings \(\mathbf{P} \in \mathbb{R}^{m \times d}\), where \(m\) is the number of prompt tokens and \(d\) is the embedding dimension. These prompt embeddings \(\mathbf{P}\) are then concatenated with hypergraph tokens \(\mathbf{G} \in \mathbb{R}^{k \times d}\), which summarise user-item interactions learned from the hypergraph encoder. To retain the position information of the tokens in the combined sequence, the positional encoding \(\mathbf{P}_{\text{pos}} \in \mathbb{R}^{(k+m) \times d}\) is applied. The fused input tokens are then passed into the downstream MoA agents, which aligns HGNN and the recommendation task representations in a shared embedding space with positional context.
\begin{equation}
\mathbf{x}_1 = \text{Concat}(\mathbf{G}, \mathbf{P}) + \mathbf{P}_{\text{pos}}
\label{eq:token_fusion}
\end{equation}.

Equation~\eqref{eq:token_fusion} fuses HGNN tokens $\mathbf{G} \in \mathbb{R}^{k \times d}$ with prompt tokens $\mathbf{P} \in \mathbb{R}^{m \times d}$ to obtain unified token embeddings $\mathbf{x}_1 \in \mathbb{R}^{(k+m) \times d}$. Concatenation preserves all token-level information from both sources while maintaining the embedding dimension $d$ consistent for downstream processing. Positional encoding $\mathbf{P}_{\text{pos}} \in \mathbb{R}^{(k+m) \times d}$ ensures that the model retains information about the relative positions of the tokens within the HGNN and prompt sequences, allowing for downstream MoA tasks.

\begin{table*}[ht]
\centering
\begin{tabular}{ccccc}
\toprule
\textbf{Statistics / Dataset} &  \textbf{Taobao} & \textbf{IJCAI} & \textbf{Tianchi} \\ 
\midrule
\# Users & 147,894 & 423,423 & 25,000 \\ 
\# Items & 99,037 & 874,328 & 500,900 \\ 
\# Interactions & 7,658,926 & 36,222,123 & 4,619,389 \\ 
Avg. Sequence Length & 51.78 & 85.54 & 184.78 \\ 
Density  & $5\times10^{-4}$ & $1\times10^{-4}$ & $4\times10^{-4}$ \\ 
Behaviour Types  & [view, fav, cart, buy] & [view, fav, cart, buy] & [view, fav, cart, buy] \\ 
Category & Dense-Medium & Sparse-Long & Dense-Long \\
\bottomrule
\end{tabular}
\caption{Statistical information of the datasets after filtering out users with fewer than five interactions.}
\label{table_datasets}
\end{table*}
\subsection{MoA Framework}
\label{subsec:moa}

The MoA module processes the fused input tokens to iteratively refine the recommendation predictions. Specifically, the MoA consists of multiple layers, each containing several agents denoted by \(A_{i,j}\), where \(i\) indexes the layer and \(j\) indexes the agent within that layer. At each layer \(i\), the input token representation \(x_i \in \mathbb{R}^{m \times d}\) is processed in parallel by all \(n\) agents \(A_{i,1}, \ldots, A_{i,n}\). Each agent \(A_{i,j}(\cdot)\) is a “frozen” LLM that produces refined token embeddings. The outputs of all agents in the layer are combined using a cross-agent attention-based aggregation operator, denoted by \(\bigoplus\). This aggregated output is then combined with the initial fused input tokens \(\mathbf{x}_1\) through a residual connection, producing the intermediate representation \(y_i\), which serves as input \(x_{i+1}\) for the next layer. 
\begin{equation}
y_i = \bigoplus_{j=1}^n A_{i,j}(x_i) + \mathbf{x}_1, \quad \text{and} \quad x_{i+1} = y_i.
\label{eq:moa_layer}
\end{equation}

\subsection{Model Training}

The training objective is to minimise the cross-entropy loss, which measures the discrepancy between predicted model predictions and ground truth. The loss function is defined as:
\begin{equation}
\mathcal{L} = -\frac{1}{N} \sum_{u=1}^N \sum_{b \in B} \sum_{i=1}^I r_{ui}^b \log \hat{r}_{ui}^b + \lambda \|\Theta\|_2^2
\end{equation}

where, \(N\) denotes the total number of users, \(I\) is the number of items, and \(B\) is the set of interaction behaviours such as \textit{view}, \textit{cart}, or \textit{purchase}. The term \(r_{ui}^b \in \{0, 1\}\) represents the ground truth label, while \(\hat{r}_{ui}^b \in (0, 1)\) is the prediction of the model. The parameter set \(\Theta\) includes trainable model components, such as user and item embeddings, HGNN weights, and MLP parameters, and \(\lambda\) controls \(L_2\) \(\|\Theta\|_2^2\) to prevent overfitting.

\begin{table*}[ht!]
\centering
\resizebox{\textwidth}{!}{%
\begin{tabular}{l l c c c c c c c}
\toprule
\textbf{Dataset} & 
\textbf{Metric} & MBHT  & PBAT & RLMRec & TRSR & KDA & TALLRec& HGLMRec \\
\midrule

\multirow{4}{*}{\centering Taobao} 
                      & HR@5($\uparrow$)  & 0.687 & 0.735 & 0.745 & 0.784 & 0.757 &\underline{0.793} & \textbf{0.832*} \\
                      & NDCG@5($\uparrow$)   & 0.590 & 0.651 & 0.657 & 0.668& 0.641 &\underline{0.685} & \textbf{0.697*} \\
                      & HR@10($\uparrow$) & 0.764 & 0.802 & 0.828 & 0.844 & 0.835 & \underline{0.849} & \textbf{0.865*} \\
                      & NDCG@10($\uparrow$) & 0.615 & 0.673 & 0.676 & 0.689 & 0.681 &  \underline{0.708} & \textbf{0.726*} \\

\midrule
\multirow{4}{*}{\centering IJCAI} 
                      & HR@5 ($\uparrow$)  & 0.774 & 0.871 & 0.882 & \underline{0.914} & 0.878 & 0.892 & \textbf{0.919*} \\
                      & NDCG@5($\uparrow$)   & 0.675 & 0.781 & 0.790 & 0.812 & 0.784 & \underline{0.797} & \textbf{0.814*} \\
                      & HR@10 ($\uparrow$) & 0.852 & \underline{0.924} & 0.913 & 0.921 & 0.906 &  0.917 & \textbf{0.932*} \\
                      & NDCG@10($\uparrow$)  & 0.701 & \underline{0.798} & 0.791 & 0.786 & 0.795 &  0.790 & \textbf{0.812*} \\
              
\midrule
\multirow{4}{*}{\centering Tianchi} 
                      & HR@5 ($\uparrow$)  & 0.718 & 0.739 & 0.764 & \underline{0.786} &  0.773 & 0.766 & \textbf{0.795*} \\
                      & NDCG@5 ($\uparrow$)  & 0.536 & 0.568 & 0.579 & \underline{0.584} & 0.581 & 0.574 & \textbf{0.606*} \\
                      & HR@10 ($\uparrow$) & 0.743 & 0.784 & 0.776 & \underline{0.797} & 0.799 & 0.783 & \textbf{0.814*} \\
                      & NDCG@10($\uparrow$) & 0.556 & 0.573 & 0.580 & 0.587 & \underline{0.601} &  0.585 & \textbf{0.613*} \\
\bottomrule
\end{tabular}
}

\caption{Performance comparison: The best performances indicated in bold show the relative improvement over the best performing baseline at 0.05 significance with paired t-test. Underlined values indicate second best performance.}
\label{c7_tab2:performance}
\end{table*}

\section{Experiments}

In our experiments, the MoA model is configured with Qwen2-7B \cite{yang2024qwen2} for the intermediate agent layers and LLaMA-3-8B \cite{dubey2024llama} for the final aggregation layer. Using different large-language models (LLMs) in the intermediate and final layers, we can leverage the strengths of each model, which in turn helps produce more coherent output, an essential factor for generating high-quality recommendations \cite{wang2024mixture}. For the \textbf{HGLM-SM} variant, we employ \textbf{LLaMA-3-70B}~\cite{dubey2024llama} and \textbf{GPT-4o}~\cite{achiam2023gpt}, selected for their reasoning abilities~\cite{zhang2024llm}. This configuration allows us to evaluate the MoA framework’s ability to generate accurate recommendations efficiently with smaller agents, while providing a meaningful baseline comparison against high-capacity models capable of handling more complex interaction patterns. Our model is implemented using PyTorch and evaluated using the Recbole framework.
 
 \paragraph{Hyperparameters.} For the \textbf{HGNN} encoder, the feature dimension is set to 128, balancing expressiveness and computational efficiency. The number of hypergraph convolution layers is set to 2 to capture high-order dependencies. We apply the AdamW optimiser for training with a learning rate of $5 \times 10^{-4}$. Warm-up steps are set to 500 to facilitate stable convergence.

\paragraph{Datasets and Baseline Models.}
For our experiments, we use three real-world e-commerce datasets that contain four types of interactions, \textit{ purchase, add-to-favourites, add-to-cart}, and \textit{view}: \textbf{Taobao} \cite{su2023personalized} , \textbf{IJCAI} \cite{su2023personalized}  and \textbf{Tianchi} \cite{elsayed2024hmar}. These datasets provide rich contextual information that enables a comprehensive evaluation of model performance in practical recommendation settings. Using these publicly available datasets also ensures that our experiments address the complexity, sparsity and diversity characteristics of real-world scenarios. The variation in the scale of the datasets further makes them valuable benchmarks for assessing models that incorporate heterogeneity and dynamics features in realistic recommendation environments. To improve the reliability of learning signals, we filter users and items with fewer than five interactions, reducing sparsity, and retaining users who exhibit multiple behavioural patterns. The summary of the datasets is shown in Table \ref{table_datasets}.
To evaluate model performance, we use Hit Ratio (HR@k) and Normalised Discounted Cumulative Gain (NDCG@k).
We compare our approach with state-of-the-art sequential methods.

\begin{itemize}

\item \textbf{MBHT:} A Multi-Behaviour Hypergraph-enhanced Transformer model which employs a multi-scale transformer to capture local user-item dependencies and a hypergraph convolution module to capture global dependencies \cite{yang2022multi}.

\item \textbf{PBAT:} A multi-behaviour  recommendation model which applies a personalised behaviour pattern generator in the representation layer and extracts dynamic behaviour patterns for sequential learning. A behaviour-aware collaboration extractor facilitates a behaviour-aware attention mechanism to incorporate behavioural and temporal impacts into collaborative transitions \cite{su2023personalized}.

\item \textbf{RLMRec} integrates representation learning with LLM to capture intricate semantic aspects of user behaviours and preferences. Incorporates auxiliary textual signals, employs LLMs for user/item profiling, and aligns the semantic space of LLMs with collaborative signals through cross-view alignment \cite{ren2024representation}. 
\item \textbf{TRSR} proposes prompt text encompassing user preference summary, recent user interactions, and candidate item information in an LLM-based recommendation, which is fine-tuned to generate recommendations \cite{zheng2024harnessing}.

 \item \textbf{KDA} uses an LLM to obtain language knowledge representations of items that are fed into a latent relation discovery module based on a discrete state variational autoencoder. Self-supervised relation discovery tasks and recommendation tasks are jointly optimised \cite{yang2024sequential}.
 \item \textbf{TALLRec:} an LLM tuning framework that structures the recommendation data as instructions to adapt to the recommendation \cite{bao2023tallrec}.
\end{itemize}


\subsection{Results and Analysis}


 Table \ref{c7_tab2:performance} presents the comparison of HGLMRec performance results against several baseline models. Generative methods RLMRec, TRSR, KDA and TALLRec show better performance compared to sequential models MBHT and PBAT, which is attributed to their generative capabilities. HGLMRec consistently outperforms baseline models in the three datasets, which is attributed to the combined benefits of the HGNN and MoA modules. The encoder provides a better capture of the user-item interactions, while the MoA module improves the ability of the model to capture diverse and complex patterns in user behaviour to enhance the recommendation. The results also show that the MoA model outperforms the models with a single LLM configuration. This performance gain is attributed to the mixture of the MoA agents, each partitioning the cognitive load to improve the accuracy of the recommendations \cite{wang2024mixture}.

In dense datasets such as Taobao and Tianchi, HGLMRec achieves substantial performance gains over baseline models. The richer interaction graphs in these datasets provide more relational and behavioural cues, which our hypergraph-based and multi-LLM agent approach can effectively exploit. Consequently, the benefits of higher-order modelling and multi-agent embedding refinement are more fully realised in these settings. In sparse datasets such as IJCAI, where much of the user–item space remains unobserved, baseline models often struggle with missing information. Here, HGLMRec still yields consistent improvements, albeit in more modest increments. However, the fact that our model does not degrade performance in these sparse settings indicates its robustness to data scarcity.

\paragraph{The MoA framework outperforms single-LLM methods.}
Table~\ref{c7_tab2:ablation} shows that MoA provides better recommendation performance compared to configurations using one large LLM. By distributing both computational and cognitive work among multiple smaller agents, MoA learns more effective representations of user-item interactions. Smaller agents also reduce computational overhead, improve scalability, and enable real-time deployment. Our evaluation confirms that the distributed workload in MoA yields better results than a single-LLM configuration.

\begin{table}[ht!]
    \centering
    \resizebox{\columnwidth}{!}{%
\begin{tabular}{lllllll}
    
    \toprule
    & \multicolumn{2}{c}{\textbf{Taobao}} & \multicolumn{2}{c}{\textbf{IJCAI}} & \multicolumn{2}{c}{\textbf{Tianchi}} \\
    \cmidrule{2-3} \cmidrule(l){4-5} \cmidrule(l){6-7}
    Model & HR@10 & N@10 & HR@10 & N@10 & HR@10 & N@10 \\   
 
    \midrule 
    w/o Hypergraph Encoder     & 0.782 & 0.645 & 0.816 & 0.762 & 0.749 & 0.568 \\
    HGLM-SM & 0.777 & 0.624 & 0.782 & 0.673 & 0.734 & 0.571 \\
  \midrule 
   HGLMRec (1 Layer)        & 0.712 & 0.608 & 0.765 & 0.614 & 0.717 & 0.538 \\
  HGLMRec (2 Layers) & 0.832 & 0.679 & 0.796 & 0.681 & 0.775 & 0.591 \\
    \hline
 HGLMRec (3 Layers) & \bfseries 0.865* & \bfseries 0.726* & \bfseries 0.932* & \bfseries 0.812* & \bfseries 0.814* & \bfseries 0.613* \\
    \bottomrule \\
\end{tabular}}
\caption{Ablation study results: Performance comparison of the full HGLMRec model and its variants.}
\label{c7_tab2:ablation}
\end{table}

\begin{figure*}
    \centering
    \includegraphics[width=1\linewidth]{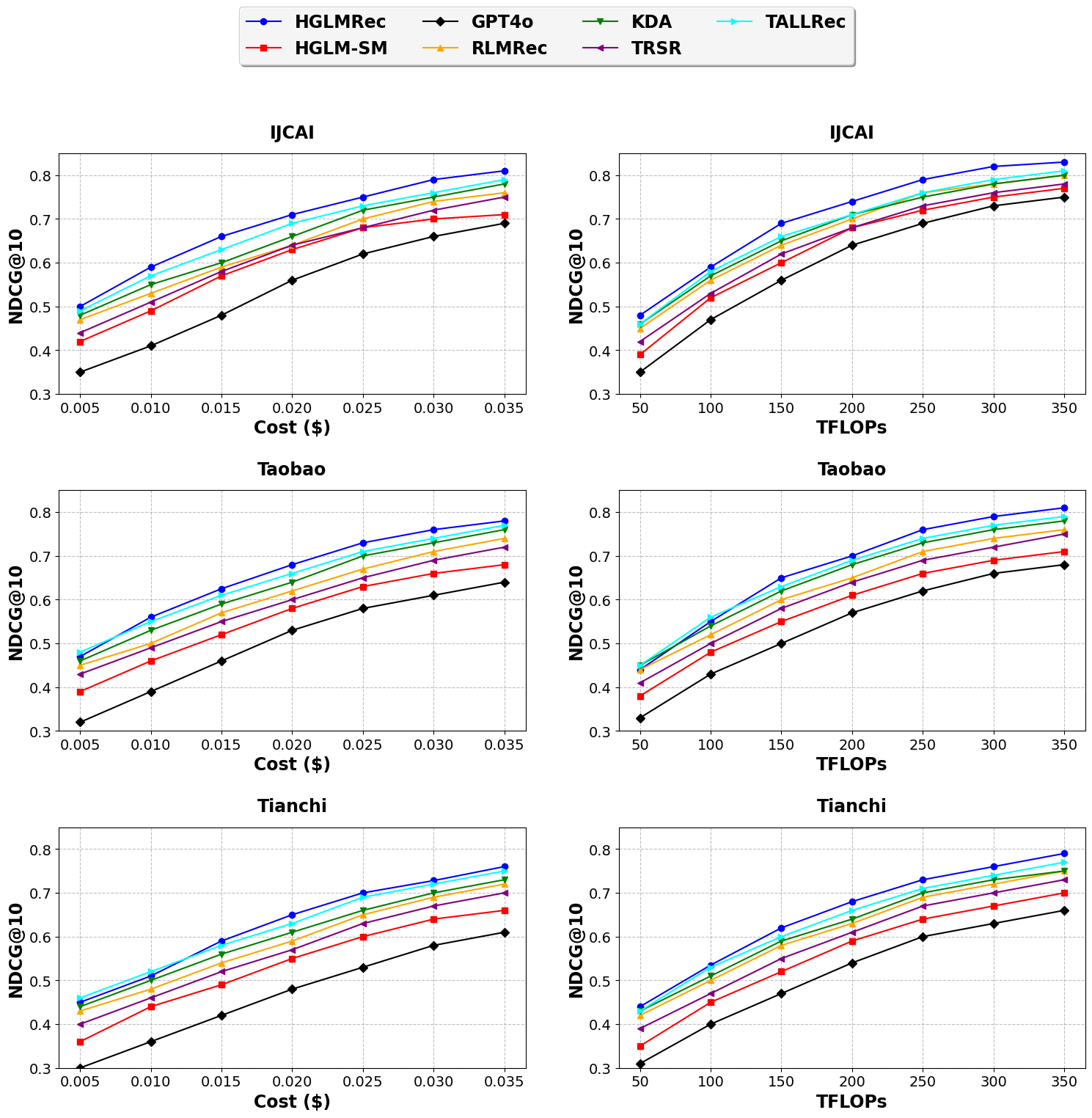}
    \Description{Cost and efficiency evaluation chart for HGLMRec.}
    \caption{Cost and efficiency evaluation of the HGLMRec model on the IJCAI dataset. The cost (in US\$) is calculated based on pricing information available on API provider websites.}
    \label{c7_fig:cost_eva}
\end{figure*}
\subsection*{Ablation Study}

The significance of the components of the HGLMRec model is evaluated by conducting experiments that determine their impact on NDCG and HR. The model variant configurations are outlined as follows: \textit{HGLM-SM:} The MoA module is replaced with a single LLM model, \textit{w/o Hypergraph Encoder:} the hypergraph encoder module is removed from the full model. The results summarised in Table \ref{c7_tab2:ablation} show that the removal of the indicated components leads to a noticeable performance degradation, highlighting the importance of each module in contributing to model performance, and the complete model consistently achieves the best results on the three datasets. In the model variant without the hypergraph encoder, performance is degraded across all datasets, which indicates that the module contributes to improving model performance. In the HGLM-SM model variant, performance also decreases compared to the HGLMRec model. 

\paragraph{Hypergraph encoder benefits.}
The encoder in HGLMRec effectively captures heterogeneous user-item interactions and facilitates richer contextual reasoning of the model, as shown by the results in Table \ref{c7_tab2:ablation}. In addition, the hypergraph representation enhances the ability of the MoA module to learn interactions between users and items, allowing a better understanding of the overall context of  user behaviour \cite{wu2022state,zhang2024llm}. The connected structure provided by the hypergraph leads to improved model learning of the user's intent, both of which are crucial for effective recommendations.

\paragraph{Performance against the number of agent layers.} The performance of HGLMRec is evaluated by varying the number of layers of LLM agents. As shown in Table \ref{c7_tab2:ablation}, the results show a clear trend: the NDCG score improves as the number of layers increases. Specifically, the model performance at 3 layers is better than variants with 1 or 2 layers. This indicates that adding more layers enhances the model's ability to capture complex patterns in user behaviour, leading to better recommendations.

\subsection*{Efficiency Evaluation} A cost comparison of HGLMRec is evaluated against other baseline models.  The cost is calculated based on the pricing information available from API providers such as OpenAI\footnote{\href{https://openai.com/api/pricing/}{https://openai.com/api/pricing.}} and Together.ai\footnote{\href{https://api.together.ai/models}{https://api.together.ai/models}}. From the results in Figure \ref{c7_fig:HGLM},  HGLMRec consistently achieves the best performance and efficiency, achieving the highest NDCG scores at the lowest computational cost. TALLRec and KDA also exhibit strong and stable performance trends, closely following HGLMRec, especially in the mid-to-high resource range. In contrast, GPT4o shows the weakest performance in all datasets and resource levels, indicating limited effectiveness despite increased computational input. HGLMRec outperforms GPT-4o and HGLM-SM in terms of both API cost and TFLOPs, achieving the highest NDCG values at every resource level, indicating superior efficiency at lower computational cost compared to methods with a single LLM configuration.

\section{Conclusion}

In this work, we introduce HGLMRec, which incorporates a hypergraph encoder to capture intricate user-item interactions while using an MoA framework. HGLMRec decouples representation learning from LLM inference and avoiding full-model fine-tuning. Experimental results show improved performance of our model compared to state-of-the-art baselines at a fraction of the computational cost of large LLM-based methods. Our approach aims not only to reduce operational costs but also to make advanced personalised recommendation more accessible in regions with limited computing infrastructure. Future work will focus on improving aspects such as fairness and reducing the latency that is introduced in the MoA layers.

\begin{acks}
This research is supported by Taighde Éireann – Research Ireland under Grant number 18/CRT/6183.
\end{acks}

\balance
\bibliographystyle{ACM-Reference-Format}
\bibliography{sample-base}


\end{document}